\documentclass[twocolumn,aps,prd,superscriptaddress,floatfix]{revtex4-1}
\usepackage{amssymb}
\usepackage{amsmath,bm}
\usepackage{graphicx}
\usepackage[normalem]{ulem}
\usepackage{bbding}
\usepackage{booktabs}
\usepackage[usenames, dvipsnames]{xcolor}
\usepackage{lipsum}
\usepackage{multirow}
\usepackage[colorlinks,
bookmarks=true,
linkcolor=blue,
urlcolor=blue,
anchorcolor=black,
citecolor=blue
]{hyperref}

\usepackage{graphicx}
\usepackage{dcolumn}
\usepackage{bm}
\usepackage{mathtools}

\usepackage{color}

\newcommand{\ud}{{\rm{d}}}

\newcommand{\ue}{{\rm{e}}}

\newcommand{\MeV}{{\ \rm{MeV}}}
\newcommand{\keV}{{\ \rm{keV}}}
\newcommand{\fm}{{\ \rm{fm}}}

\newcommand{\kpc}{{\ \rm{kpc}}}
\newcommand{\erg}{{\ \rm{erg}}}

\definecolor{Green}{RGB}{0, 128, 0}
\graphicspath{{./Figures/}}

\renewcommand\sout{\bgroup \color{red} \ULdepth=-.5ex \ULset}

\begin{document}

\title{Supernova Neutrinos as a Precise Probe of Nuclear Neutron Skin}

\author{Xu-Run Huang}
\affiliation{School of Physics and Astronomy, Shanghai Key Laboratory for Particle Physics and Cosmology, and Key Laboratory for Particle Astrophysics and Cosmology (MOE), Shanghai Jiao Tong University, Shanghai 200240, China}

\author{Lie-Wen Chen}
\thanks{Corresponding author}
\email{lwchen@sjtu.edu.cn}
\affiliation{School of Physics and Astronomy, Shanghai Key Laboratory for Particle Physics and Cosmology, and Key Laboratory for Particle Astrophysics and Cosmology (MOE), Shanghai Jiao Tong University, Shanghai 200240, China}

\date{\today}

\begin{abstract}

A precise and model-independent determination of the neutron distribution radius $R_{\rm n}$ and thus the neutron skin thickness $R_{\rm skin}$ of atomic nuclei is of fundamental importance in nuclear physics, particle physics and astrophysics but remains a big challenge in terrestrial labs.
We argue that
the nearby core-collapse supernova (CCSN) in our Galaxy may render a neutrino flux with unprecedentedly high luminosity, offering perfect opportunity to determine the $R_{\rm n}$ and $R_{\rm skin}$ through the coherent elastic neutrino-nucleus scattering (CE$\nu$NS).
We evaluate
the potential of determining the $R_{\rm n}$ of lead (Pb) via CE$\nu$NS with the nearby CCSN neutrinos in the RES-NOVA project which is designed to hunt CCSN neutrinos using an array of archaeological Pb based cryogenic detectors.
We find that
an ultimate precision of $\sim 0.1 \%$ for the $R_{\rm n}$ ($\sim 0.006$ fm for the $R_{\rm skin}$) of Pb can be achieved via RES-NOVA in the most optimistic case that the CCSN explosion were to occur at a distance of $\sim 1 \kpc$ from the Earth.

\end{abstract}

\maketitle

\section{\label{sec:Int}Introduction}
Neutrons are expected to be distributed more extensively than protons in heavy neutron-rich nuclei, forming a neutron skin which is featured quantitatively by the skin thickness $R_{\rm skin} =R_n - R_p$ where $R_{n}$ and $R_{p}$ are the (point) neutron and proton rms radii of the nucleus, respectively.
Theoretically, it has been established that the $R_{\rm skin}$ provides an ideal probe for the density dependence of
the symmetry energy
$E_{\rm sym}(\rho)$~\cite{Brown:2000pd,Typel:2001lcw,Furnstahl:2001un,Yoshida:2004mv,Chen:2005ti,Centelles:2008vu,Chen:2010qx,Reinhard:2010wz,Roca-Maza:2011qcr,Agrawal:2012pq,Zhang:2013wna,Mondal:2016bls,Raduta:2017cma,Newton:2020jwn,Lynch:2021xkq},
which quantifies the isospin dependent part of the equation of state (EOS) for isospin asymmetric nuclear
matter and plays a critical role in many issues of nuclear physics and astrophysics~\cite{Danielewicz:2002pu,Lattimer:2004pg,Steiner:2004fi,Baran:2004ih,Li:2008gp,Horowitz:2014bja,Gandolfi:2015jma,Ozel:2016oaf,Baldo:2016jhp,Roca-Maza:2018ujj,Drischler:2021kxf,Li:2021thg}.

Experimentally,
while the $R_{p}$ can be precisely inferred from its corresponding charge rms radius $R_{\rm ch}$ which has been measured precisely via electromagnetic processes~\cite{Fricke:1995zz,Angeli:2013epw}, the $R_{n}$ remains elusive since it is usually
determined from strong processes, generally involving in model dependence (see,
e.g., Ref.~\cite{Thiel:2019tkm}).
A clean approach to determine the $R_{n}$ is to measure
the parity-violating asymmetry $A_{\rm PV}$ in the elastic
scattering of polarized electrons from the nucleus since the $A_{\rm PV}$ is particularly sensitive to the neutron distribution due to its large weak charge compared to the tiny one of the
proton~\cite{Donnelly:1989qs,Horowitz:1999fk}.
Following this strategy,
the $^{208}$Pb radius experiment (PREX-2)~\cite{PREX:2021umo} and $^{48}$Ca radius experiment (CREX)~\cite{CREX:2022kgg} recently reported the determination of the $R_n$ with a precision of $\sim 1\%$, i.e., $R^{208}_{\rm skin} = 0.283\pm0.071~\rm{fm}$ for $^{208}$Pb~\cite{PREX:2021umo} and $R^{48}_{\rm skin} = 0.121\pm0.026(\rm exp)\pm 0.024(\rm model)$~fm for $^{48}$Ca ($1\sigma$ uncertainty).
Very remarkably,
analyses within modern energy density functionals~\cite{Reinhard:2022inh,Yuksel:2022umn,Zhang:2022bni}
conclude a tension between the CREX and PREX-2 results, with the former favoring a very soft $E_{\rm sym}(\rho)$ while the latter a very stiff one, calling for further critical theoretical and experimental investigations.
Especially, the Bayesian analysis~\cite{Zhang:2022bni} suggests that a higher precision for the $R_n$ of $^{208}$Pb is of particular importance to address this issue.
The Mainz Radius Experiment (MREX)~\cite{Becker:2018ggl}
is expected to shrink the uncertainty by a factor of two with a precision of $0.5 \%$ (or $\pm 0.03 \fm$) for the $R_n$ of $^{208}$Pb, but the experiment's start time is still largely uncertain~\cite{Middleton:2021pt}.

Another clean and model-independent way to extract the $R_{\rm skin}$ is through the coherent elastic neutrino-nucleus scattering (CE$\nu$NS)~\cite{Freedman:1973yd,Freedman:1977xn}, which was firstly observed by the COHERENT Collaboration via a CsI detector with the
neutrino beam from the Spallation Neutron Source at Oak Ridge National Laboratory~\cite{COHERENT:2017ipa}.
Based on the COHERENT data, the $R_{\rm skin}$ of CsI has been extracted~\cite{Cadeddu:2017etk,Huang:2019ene}  but the
uncertainty is too large to claim a determination, due to the low statistics of CE$\nu$NS events.
In nature,
the nearby core-collapse supernova (CCSN) may render a neutrino flux with unprecedentedly high luminosity,
which provides an excellent chance to explore CE$\nu$NS.
Indeed, detecting the next galactic SN neutrinos
has received much attention both from large neutrino observatories and modern dark matter experiments~\cite{Scholberg:2012id,DARWIN:2016hyl,JUNO:2015zny,DUNE:2020zfm,DarkSide20k:2020ymr,Hyper-Kamiokande:2021frf,Pattavina:2020cqc,RES-NOVA:2021gqp}. One of the most powerful projects
is the RES-NOVA experiment which will hunt CCSN neutrinos via
CE$\nu$NS by adopting an archaeological Pb based cryogenic detector~\cite{Pattavina:2020cqc,RES-NOVA:2021gqp}.
One merit of RES-NOVA is that using CE$\nu$NS as its detection channel allows a flavor-blind neutrino measurement and thus avoids the uncertainties from the neutrino oscillation.
The other merit is that archaeological Pb ensures the large CE$\nu$NS cross section and the ultra-low levels of background, literally guaranteeing a high statistics.

In this work,
we demonstrate that the very configuration of the RES-NOVA experiment provides an ideal site to determine
the $R_{\rm n}$ of Pb, and
an ultimate precision of $\sim 0.1 \%$ for the $R_{\rm n}$ ($\sim 0.006$ fm for the $R_{\rm skin}$) of Pb can be achieved in the most optimistic case that the galactic CCSN would explode at a distance of $\sim 1 \kpc$ from the Earth.
Even with a CCSN at 5 kpc, our present approach can still achieve a precision better than that from PREX-2.

The paper is organized as follows. In Section~\ref{sec:FluxSN}, we give a brief description of the supernova neutrinos.
In Section~\ref{sec:RES-NOVA}, we discuss the prospects of the neutrino detection in RES-NOVA experiment.
In Section~\ref{sec:NSkin}, the results on the neutron skin thickness sensitivity are presented and discussed.
The conclusions are given in Section~\ref{sec:Sum}.

\section{\label{sec:FluxSN}Supernova neutrinos}
The detailed knowledge of a SN neutrino flux is still missing in experiments since we have only observed two dozen neutrino events from the SN 1987A~\cite{Bionta:1987qt,Kamiokande-II:1987idp}. However, after three decades, current neutrino experiments have stepped into an era with unprecedented accuracy. The robust reconstruction of SN neutrino spectra with multiple detectors has been investigated~\cite{Minakata:2008nc,Dasgupta:2011wg,Lu:2016ipr,Nikrant:2017nya,GalloRosso:2017mdz,Li:2017dbg,Li:2019qxi,GalloRosso:2020qqa,Nagakura:2020bbw} and an accurate measurement is promising for a nearby SN (e.g., $< 5 \kpc$).
Furthermore,
modern SN simulations have achieved a tremendous progress in unveiling the mysteries of SN phenomena~\cite{Janka:2016fox,Muller:2016izw,Just:2018djz,OConnor:2018sti,Nagakura:2020qhb}.
Based on current understanding, the spectral shape of CCSN neutrino fluxes for each flavor can be well approximated by a pinched thermal distribution~\cite{Keil:2002in,Tamborra:2012ac}
\begin{equation}
\label{eq:vSpectrum}
    f_{\nu}(E_\nu) = A \left(\dfrac{E_\nu}{\left< E_\nu \right>}\right)^\alpha \rm{exp}\left[ - (\alpha +1) \dfrac{E_\nu}{\left< E_\nu \right>} \right].
\end{equation}
Here, $E_\nu$ and $\left< E_\nu \right>$ are the neutrino energy and the averaged energy, $\alpha$ describes the amount of spectral pinching, and $ A = \dfrac{(\alpha +1)^{\alpha + 1}}{\left< E_\nu \right> \Gamma(\alpha + 1)}$ is the normalization constant, where $\Gamma$ is the gamma function. So the neutrino fluence per flavor on the Earth from a CCSN at a distance $d$ can be obtained as
\begin{equation}
\label{eq:vFluence}
    \Phi(E_\nu) = \dfrac{1}{4 \pi d^2} \dfrac{E_\nu^{\rm tot}}{\left< E_\nu \right>} f_{\nu}(E_\nu),
\end{equation}
where $E_\nu^{\rm tot}$ denotes the total emitted energy per flavor.
In a real CCSN explosion,
both the amounts and spectra of the emitted neutrinos change with time as the star evolves into different stages. However, to our goal, we only need information of total neutrino emission.
Therefore, we adopt here the time-integrated neutrino emission parameters from a typical long-term axisymmetric CCSN simulation, which can be found in Table I of Ref.~\cite{Nikrant:2017nya}.
Note that although the neutrino emission of a CCSN also depends on the details of the transient, e.g., the progenitor mass, compactness, explosion dynamics, etc., it has a rough profile of $\left< E_\nu \right> \sim 10 \MeV$, $2 < \alpha < 4$ and $E^{\rm tot} \sim 10^{53} \erg$. Nevertheless, the accurate information can be extracted from various detection data once a nearby CCSN explosion occurs.

\section{\label{sec:RES-NOVA}Detection prospects in RES-NOVA}
To explore the potential of determining the $R_{\rm n}$ of Pb with RES-NOVA, we consider the RN-3 configuration in Table I of Ref.~\cite{Pattavina:2020cqc} which has a detector mass of 465 t and an energy threshold of $1 \keV$. The absorber with pure Pb is also adopted.
For the detection channel,
the differential cross section in the standard model has the form:
\begin{eqnarray}
\dfrac{\ud \sigma}{\ud T}(E_\nu,T) = \dfrac{G_F^2 M}{4 \pi} Q_W^2 F_W^2(q) \Big[1 - \dfrac{T}{E_\nu} - \dfrac{MT}{2 E_\nu^2}  \Big], \label{eq:croSection}
\end{eqnarray}
where $G_F$ is the Fermi coupling constant; $M$ denotes the mass of the target nucleus with $N(Z)$ neutrons (protons); $Q_W$ is the weak charge and $F_W(q)$ is the weak form factor; $E_\nu$ and $T$ represent the neutrino energy and the kinetic recoil energy of the nucleus, respectively; and the momentum transfer $q$ is given by $q^2 \simeq 2 M T$.
Note Eq.~(\ref{eq:croSection}) is for a nucleus with spin-0 and the result for a spin-1/2 target (i.e., $^{207}$Pb in our case) will gain a tiny correction~\cite{Lindner:2016wff} which is neglected in this work.

The weak charge $Q_W$ can be obtained as
\begin{equation}
\label{weakCharge}
	Q_W = \int \mathrm{d}^3 r \rho_W(r) = N q_n + Z q_p ,
\end{equation}
where $\rho_W(r)$ is the weak charge density.
At tree level, the nucleon weak charges are $q_n = q_n^0 = 2 g_V^n$ and $q_p = q_p^0 = 2 g_V^p$, where the neutron (proton) vector coupling is defined as $g_V^n = - \frac{1}{2}$ ($g_V^p = \frac{1}{2} - 2 \sin^2 \theta_W$) with the low-energy weak mixing angle $\sin^2 \theta_W = 0.23857(5)$~\cite{Kumar:2013yoa,ParticleDataGroup:2018ovx}.
In the present work, we adopt the values $q_n = -0.9878$ and $q_p = 0.0721$ to include radiative corrections~\cite{Horowitz:2012tj}.
The weak form factor $F_W(q)$ is expressed as
\begin{equation}
	F_W(q) = \dfrac{1}{Q_W} \int \mathrm{d}^3 r \dfrac{\sin q r}{q r} \rho_W(r).
\end{equation}
Here we use the Helm parametrization for the $F_W(q)$~\cite{Helm:1956zz,Piekarewicz:2016vbn}, which has been proven to be very successful for analyzing electron scattering form factors~\cite{PhysRev.163.927,Raphael:1970yd}. The $F_W(q)$ is then expressed as
\begin{equation}
\label{eq:helmFF}
F_W(q) = 3 \dfrac{j_1(q R_0)}{q R_0} \ue^{- q^2 s^2 /2},
\end{equation}
where $j_1(x) = \sin(x)/x^2 - \cos(x)/x$ is the spherical Bessel function of order one, $R_0$ is the diffraction radius and $s$ quantifies the surface thickness. The rms radius $R_W$ of weak charge density can then be obtained as
\begin{equation}
\label{eq:rmsRadius}
R_W^2 = \int \mathrm{d}^3 r \dfrac{r^2 \rho_W(r)}{Q_W} = \dfrac{3}{5} R_0^2 + 3 s^2.
\end{equation}
We use $s = 1.02 \fm$ following the discussion in Ref.~\cite{Horowitz:2012tj}.
The $R_n$ and $R_p$ are related to $R_W$ and $R_{\rm ch}$ with the following relations~\cite{Ong:2010gf,Horowitz:2012tj},
\begin{equation}
\label{eq:RpRch}
	R_p^2 = R^2_{\rm ch} - \langle r_p^2 \rangle - \dfrac{N}{Z} \langle r_n^2 \rangle
\end{equation}
and
\begin{equation}
\label{eq:RnRw}
	R_n^2 = \dfrac{Q_W}{q_n N} R_W^2 - \dfrac{q_p Z}{q_n N} R_{\rm ch}^2 - \langle r_p^2 \rangle - \dfrac{Z}{N} \langle r_n^2 \rangle + \dfrac{Z+N}{q_n N} \langle r_s^2 \rangle.
\end{equation}
Here $\langle r_p^2 \rangle^{1/2} = 0.8414(19) \fm$~\cite{Hammer:2019uab} is the charge radius of a proton and $\langle r_n^2 \rangle = -0.1161(22) \fm^2$~\cite{ParticleDataGroup:2018ovx} is that of a neutron, the squared strangeness radius of nucleon is taken to be $\langle r_s^2 \rangle = -0.0054(16) \fm^2$ according to Lattice QCD calculations~\cite{Horowitz:2018yxh,Green:2015wqa}.
Note that the contributions of the Darwin-Foldy term and the spin-orbit current are neglected here since both of them are quite small and will not affect our conclusions on the relative precision evaluation of the $R_n$ determination for Pb.

\begin{table}[b]
\caption{\label{tab:Pb}%
Charge radii $R_{\rm ch}$~\cite{Angeli:2013epw}, binding energies per nucleon $E_B/A$~\cite{Wang:2021xhn} and abundance $Y_A$~\cite{Meija:2016pac} of Pb isotopes.}
\begin{ruledtabular}
\begin{tabular}{cccc}
\textrm{Isotopes}&
\textrm{$R_{\rm ch}$ (\rm{fm})}&
\textrm{$E_B/A$ ($\rm{MeV}$)}&
\textrm{$Y_A$}\\
\colrule
204 & 5.4803(14) & 7.87993 & 0.014(6)\\
206 & 5.4902(14) & 7.87536 & 0.241(30)\\
207 & 5.4943(14) & 7.86987 & 0.221(50)\\
208 & 5.5012(13) & 7.86745 & 0.524(70)\\
\end{tabular}
\end{ruledtabular}
\end{table}

The archaeological Pb crystal in RES-NOVA is mainly composed of four isotopes, i.e., $^{204,206,207,208}$Pb. The charge rms radius $R_{\rm ch}$,
the binding energies per nucleon $E_B/A$ and the natural abundance $Y_A$ of these isotopes can be found in Table~\ref{tab:Pb}.
Since it is impossible to
distinguish the CE$\nu$NS events from different isotopes in RES-NOVA, what we can extract from such detection is the averaged $R_{\rm skin}$  of Pb and that is what we really mean for the $R_{\rm skin}$  of Pb in this work.
The mass of a nucleus is defined as $M = N \times m_n + Z \times m_p -E_B$ where $E_B$ is the binding energy and $m_{n(p)}$ is the rest mass of neutrons(protons).

We first assume a constant $100 \%$ acceptance efficiency in the detector, and the expected event counts can then be obtained as
\begin{equation}
\label{eq:events}
    \dfrac{\ud N}{\ud T} = \sum Y_A N_t \int_{E_{\rm min}} \ud E_\nu \ \Phi(E_\nu) \ \dfrac{\ud \sigma}{\ud T}(E_\nu, T).
\end{equation}
Here, $N_t = N_A m_{\rm det} / M_{\rm{Pb}}$ is the number of nuclei in the crystals with $N_A$ being the Avogadro constant, $m_{\rm det}$ the detector mass and $M_{\rm{Pb}} = 0.2072 \ \rm{kg/mol}$ the molar mass of Pb. Strictly speaking, the value of $E_{\rm min}$ depends on $T$ due to the relation: $T_{\rm max} = 2 E^2_\nu / (M + 2 E_\nu)$. We adopt $T_{\rm max} \simeq 2 E^2_\nu / M$ since $M \gg E_\nu$ in this scattering. The final count will sum over both the isotopes and neutrino flavors, and integrate over the corresponding energy bin.

The final result will vitally depend on the distance of the SN.
On the one hand,
the target SN cannot be too far since the event rate has an inverse quadratic dependence on the distance, as shown in Eq.~(\ref{eq:vFluence}) and Eq.~(\ref{eq:events}). Especially, the relative high energy part ($T > 10 \keV$) has rather low event rate and even gets hidden under the background for $d = 10 \kpc$~\cite{Pattavina:2020cqc}. However, recent study shows there exists shock acceleration in SN to create more high energy heavy flavor neutrinos~\cite{Nagakura:2020gls}. Note that the nucleon distribution radii exhibit stronger sensitivity to the form factor at higher momentum transfer in this low energy range. We thus choose $d = 5 \kpc$ for a typically far distance SN target.
On the other hand,
a very nearby SN will lead to a neutrino flux intense enough to cause signal pile-up in the detector. This phenomenon has been studied recently by the RES-NOVA Collaboration, and the results show that the pile-up probability will decrease to almost zero for $d \gtrsim 1 \kpc$ ~\cite{RES-NOVA:2021gqp}. Therefore, $d = 1 \kpc$ turns out to be an optimal choice.

\begin{figure}[t!]
	\centering
	\includegraphics[width=0.95\columnwidth]{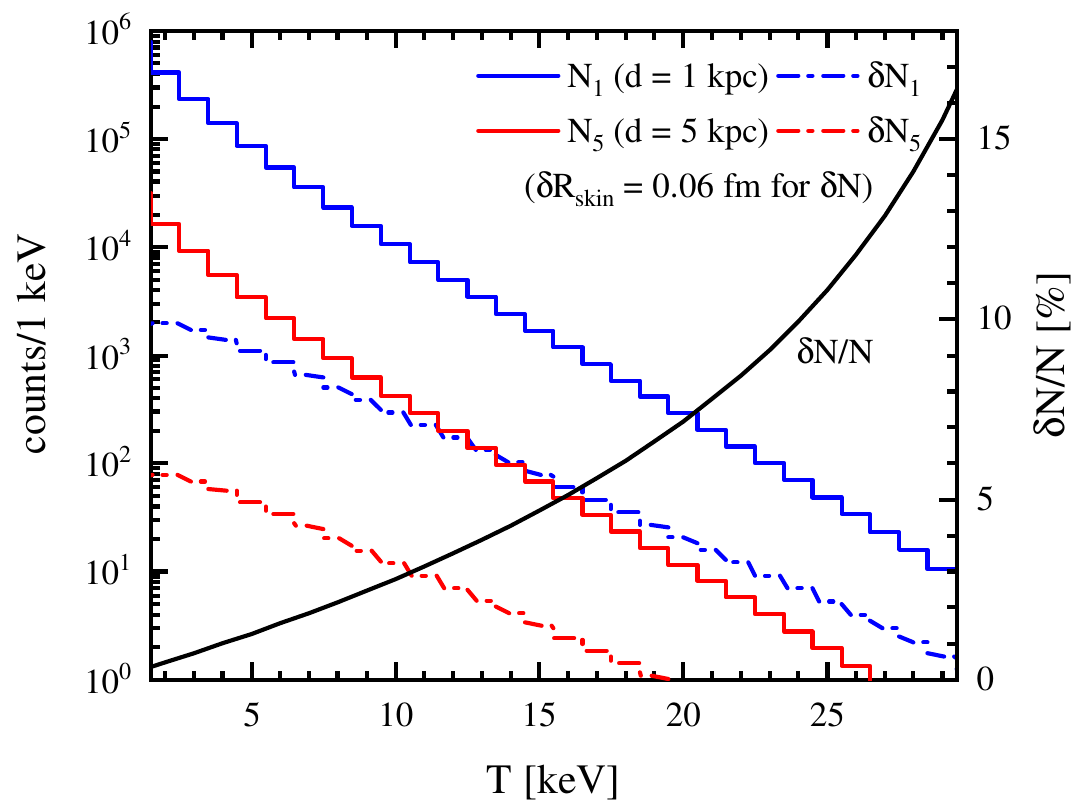}
	\caption{The predicted event counts per 1 keV versus nuclear kinetic recoil energy $T$ at the RES-NOVA detector for a SN at $1 \kpc$ ($N_1$, blue solid line) and $5 \kpc$ ($N_5$, red solid line), using the PREX-2 result as the averaged $R_{\rm skin}$ of Pb. The dashed lines ($\delta N_1$ and $\delta N_5$) show the change amplitude of counts with a $0.06 \fm$ variation of the neutron skin thickness. The relative variation $\delta N/N$ of the counts is given for both SN distances by the black line (right axis).}
	\label{fig:events}
\end{figure}

The expected results are shown in Fig.~\ref{fig:events}. We adopt an energy bin of $1 \keV$, which is allowable since the energy resolution of RES-NOVA is expected to be $0.2 \keV$. The statistics is promising for both cases of $d = 1 \kpc$ and $d = 5 \kpc$.
In particular,
the count per bin ranges from $10^3$ to $10^5$ in the recoil energy range ($3-19 \keV$) for a SN at $1 \kpc$, while it becomes approximately one order of magnitude smaller for $d = 5 \kpc$.
We also estimate
the sensitivity to the variation of the $R_{\rm skin}$.
As an example, we plot the count  change for a $R_{\rm skin}$ variation of $0.06 \fm$ ($\sim 1 \%$).
The modifications on counts for $d = 1(5) \kpc$ stay within about $10^1(1) - 10^3(10^2)$ depending on the recoil energy $T$.
To elucidate this effect more clearly, the relative difference $\delta N/N$ is also shown by the black line. As anticipated, the $\delta N/N$ only depends on the recoil energy for a certain variation of the $R_{\rm skin}$ and increases with the recoil energy.
This is because it directly quantifies the modifications on the form factor.
On the other hand, one sees that both the expected counts and its variation $\delta N$ decrease exponentially with the recoil energy. This is due to the roughly exponential decay of SN neutrino flux as a function of neutrino energy at higher energies as shown in Eq.~(\ref{eq:vSpectrum}).
As a result,
considering both the sensitivity and event statistics, the bins around the center ($T\sim 15$~keV) are more suitable to measure the $R_{\rm skin}$.

\section{\label{sec:NSkin}Neutron skin thickness sensitivity}
In order to carry out a statistical evaluation on the precision of the $R_n$ determination, we follow Ref.~\cite{Canas:2018rng} and adopt the following chi-squares function:
\begin{equation}
\label{eq:chiSquare}
    \chi^2 = \sum_{\rm bins} \dfrac{(N_{\rm exp} - N_{\rm th})^2}{\sigma_{\rm stat}^2 + \sigma^2_{\rm syst}}.
\end{equation}
Here, the ``experimental" data $N_{\rm exp}$ is taken as the expected counts with $R_{\rm skin} = 0.283 \fm$ and the theoretical counts $N_{\rm th}$ will vary according to the $R_{\rm skin}$ value.
Except for the statistical uncertainty $\sigma_{\rm stat} = \sqrt{N_{\rm exp}}$, we also introduce an effective systematic error $\sigma_{\rm syst} = p N_{\rm th}/100$ to quantify the possible uncertainties originating from the future RES-NOVA detector, the extracted SN neutrino spectra from other detection, and so on,
with $p$ representing the percentage of the future systematic error.
The sum in Eq.~(\ref{eq:chiSquare}) only runs over the energy bins around the center part at $T\sim 15$~keV (see the following for the detailed values).
For bins with lower energy, they have higher statistics but much less sensitivity to the neutron form factor due to their almost full coherence. Moreover, the previous assumption of an energy-independent $100 \%$ acceptance efficiency is more likely to become unacceptable in the energy bins near the threshold of detector. We thus drop the first 3 bins above the energy threshold and choose the minimum recoil energy to be $T_{\rm min} = 4.5 \keV$, which corresponds to a minimum neutrino energy of $E_{\rm min} \simeq 20.9 \MeV$. That means that only the high energy SN neutrinos contribute to this analysis.
In contrast, bins in higher energy region show better sensitivity but poor statistics. We adopt $T_{\rm max} = 29.5(20.5) \keV$
[corresponding to a minimum neutrino energy of $E_{\rm min} \simeq 53.4(44.6) \MeV$]
for $d = 1(5) \kpc$ just to ensure that every bins have a reasonable event number (i.e. $N > 10$).

\begin{figure}[t!]
 \centering
 \includegraphics[width=0.95\columnwidth]{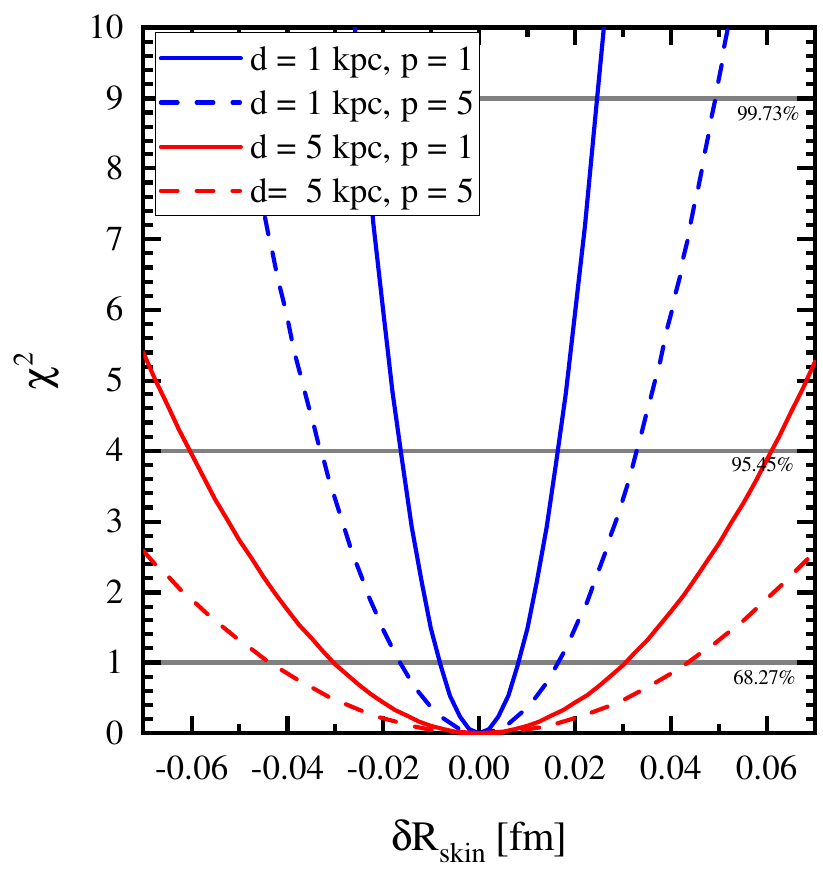}
 \caption{The expected sensitivity to the variation of neutron skin thickness $\delta R_{\rm{skin}}$ for a SN at $1 \kpc$ (blue lines) and $5 \kpc$ (red lines) with $1 \%$ (solid linces) and $5 \%$ (dashed lines) systematic error.}
\label{fig:sensitivity}
\end{figure}

The resulting $\chi^2$ as a function of the variation $\delta R_{\rm{skin}}$ for the neutron skin thickness is shown in Fig.~\ref{fig:sensitivity} where the future systematic error is presumed to be $1 \%$ or $5 \%$.
In particular, the $1\sigma$ uncertainty for a $1(5) \%$ systematic error is $\pm 0.008(0.016) \fm$ if the SN is located at $d = 1 \kpc$, and $\pm 0.030(0.044) \fm$ at $d = 5 \kpc$.
It is remarkable that our approach can achieve an ultimate precision of $\sim 0.1 \%$ in the optimal case of $d = 1 \kpc$, even much higher than the expected precision of the future MREX~\cite{Becker:2018ggl}.

\begin{figure}[t!]
	\centering
	\includegraphics[width=0.95\columnwidth]{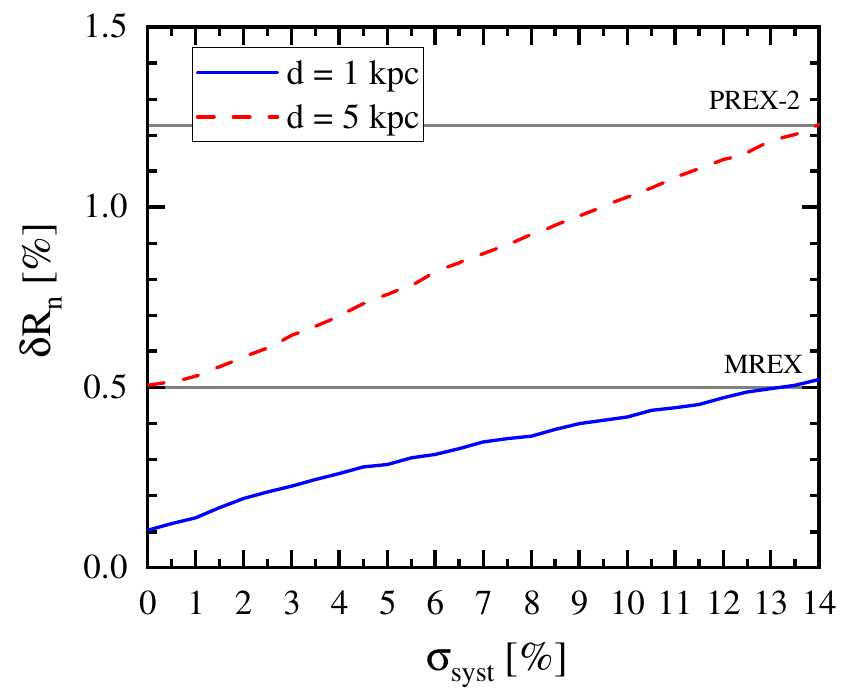}
	\caption{The expected precision of the averaged neutron radius $R_n$ (in percent) as a function of the systematic uncertainty for a $1 \kpc$ (blue solid line) and $5 \kpc$ (red dashed line) SN. The results from PREX-2~\cite{PREX:2021umo} and the future MREX~\cite{Becker:2018ggl} are also included for comparison.}
	\label{fig:SystUncertainty}
\end{figure}

The systematic error $\sigma_{\rm syst}$ is of great importance in future experiments and observations, and our current knowledge on $\sigma_{\rm syst}$ is rather insufficient. To see more clearly how the precision depends on the $\sigma_{\rm syst}$,
we plot in Fig.~\ref{fig:SystUncertainty} the expected $1\sigma$ precision (in percentage) for the $R_n$ determination as a function of the $\sigma_{\rm syst}$.
It is seen that the $\sigma_{\rm syst}$ dependence is nearly unaffected by the SN distance and is almost linear.
Generally, the closer the SN is located, the better its neutrino flux can be measured, and thus the smaller $\sigma_{\rm syst}$ from the spectra will be achieved.
In particular, the determination of the $R_n$ can achieve a precision better than that of MREX for a nearby SN at $d \simeq 1 \kpc$ as long as $\sigma_{\rm syst} \lesssim 13 \%$ as shown in Fig.~\ref{fig:SystUncertainty}.
Even under a worse condition at $d \simeq 5 \kpc$, one can still anticipate a precision better than that of PREX-2.
At this point,
we would like to emphasize that
the nearby presupernova stars are not too rare in our Galaxy.
For example, a list of 31 candidates within 1 kpc, including the famous
Betelgeuse, can be found in Ref.~\cite{Mukhopadhyay:2020ubs}.
Accordingly, more than $\sim 750$ candidates are expected to exist in $1\sim 5$ kpc assuming the presupernova stars are uniformly distributed around the Earth in the Milky Way Disk.
In addition, assuming the CCSN rate is about 2 per 100 years in our Galaxy and the rate is further assumed to be unform for a rough estimate,
one then obtains a rate of $2\times (5~\rm kpc/15~\rm kpc)^2 \approx 0.2$ per 100 years within 5 kpc.
On the other hand,
it is very interesting to note that there are totally six galactic SNe which have been recorded so far since 1000 A.D., i.e.,
Lupus at 2.2 kpc in 1006 (SN~1006), Crab at 2.0 kpc in 1054 (SN~1054), 3C 58 at 2.6 kpc in 1181 (SN~1181), Tycho at 2.4 kpc in 1572 (SN~1572), Kepler at 4.2 kpc in 1604  (SN~1604), and Cas A at 2.92 kpc in 1680 (SN~1680),
and they all occurred within $5$ kpc from the Earth~\cite{The:2006iu}.
Among the six galactic SNe,
four of them, namely, Crab (SN~1054), 3C 58 (SN~1181), Kepler (SN1604) and Cas A (SN~1680) were considered to be created by CCSNe~\cite{The:2006iu}.
However, it should be noted that more recent studies indicate the Kepler (SN1604) seems to be now generally regarded to have been a type Ia supernova, although a surviving donor has not been detected (see, e.g., Ref.~\cite{2017hsn..book..139V}).
Therefore, there are at least three recorded CCSNe (i.e., SN~1054, SN~1181 and SN~1680) so far within $5$ kpc from the Earth since 1000 A.D..
In particular,
the most recent recorded CCSN, i.e., Cas A, occurred more than 340 years ago (in 1680)~\cite{The:2006iu}.
Based on these observations, we conclude that while it is hard to predict precisely when the next nearby CCSN would occur, one may still expect optimistically that the nearby ($\lesssim 5$~kpc) CCSN seems to be imminent.

Furthermore, it is instructive to have a discussion on the measurement of $\nu$ spectra from a CCSN since it contributes to $\sigma_{\rm syst}$.
For a flavor-blind measurement of $\nu$ spectra for our present motivation,
one only needs to know the configuration of total $\nu$ flux ($\nu^0$) when they are freshly produced in the CCSN.
At this initial stage,
heavy flavor neutrinos ($\nu^0_x$ and $\bar{\nu}^0_x$ with $x=\mu,\tau$) contribute to the total $\nu^0$ flux by $ \sim 2/3$ with the neutrinos and their anti-neutrinos having equal fraction ($\nu^0_x = \bar{\nu}^0_x$), while $\nu^0_e$ and $\bar{\nu}^0_e$ contribute to the rest $ \sim 1/3$.
However, the neutrino flavor conversions will occur during the neutrinos propagate before they eventually reach terrestrial detectors.
Assuming adiabatic conversion in the SN, in the case of the normal mass ordering, the observed luminosity of a neutrino species $L_{\nu_i}^{\rm obs}$ is~\cite{Scholberg:2017czd}
\begin{eqnarray}
L_{\nu_e}^{\rm obs} &=& L_{\nu_x^0}, \\
L_{\bar{\nu}_e}^{\rm obs} &=& {\rm cos}^2\theta_{12} L_{\bar{\nu}_e^0} + {\rm sin}^2\theta_{12} L_{\bar{\nu}_x^0},
\end{eqnarray}
where $\theta_{12}$ is the mixing angle between mass eigenstates $\nu_1$ and $\nu_2$.
In the case of the inverted mass ordering, the observed luminosity of a neutrino species $L_{\nu_i}^{\rm obs}$ is~\cite{Scholberg:2017czd}
\begin{eqnarray}
L_{\nu_e}^{\rm obs} &=& {\rm sin}^2\theta_{12} L_{\nu_e^0} + {\rm cos}^2\theta_{12} L_{\nu_x^0}, \\
L_{\bar{\nu}_e}^{\rm obs} &=& L_{\bar{\nu}_x^0}.
\end{eqnarray}
Therefore,
after the neutrino flavor conversions,
the $\nu^0_x$ or $\bar{\nu}^0_x$ spectra can be determined from the signals of $\nu_e$ (e.g., DUNE~\cite{DUNE:2020zfm}) or $\bar{\nu}_e$ (e.g., Hyper-K~\cite{Hyper-Kamiokande:2021frf}), respectively, for a given neutrino mass ordering (i.e., $\nu_e = \nu^0_x$ for normal mass ordering while $\bar{\nu}_e = \bar{\nu}^0_x$ for inverted mass ordering).
On the other hand, after the neutrino flavor conversions,
for normal (inverted) mass ordering,
the spectra of $\bar{\nu}^0_e$ ($\nu^0_e$) (roughly $\sim 1/6$ in $\nu^0$) can be extracted from the $\bar{\nu}_e$ ($\nu_e$) signals
while information of $\nu^0_e$ ($\bar{\nu}^0_e$) is only carried by $\nu_x$ ($\bar{\nu}_x$) signals which can be measured by dark matter detectors~\cite{DarkSide20k:2020ymr} or other neutral current detectors (e.g., JUNO~\cite{JUNO:2015zny}).

Finally, we would like to mention that the advanced ab initio approaches using nuclear
forces from chiral effective field theory can now describe the properties of heavy nuclei such as $^{208}$Pb~\cite{Hu:2021trw}.
In particular, the ab initio calculations predict $R^{48}_{\rm skin} = 0.141 - 0.187$~fm and $R^{208}_{\rm skin} = 0.139 - 0.200$~fm~\cite{Hu:2021trw}, consistent with the CREX result of $R^{48}_{\rm skin} = 0.121\pm0.026(\rm exp)\pm 0.024(\rm model)$~fm~\cite{CREX:2022kgg} but exhibiting a mild tension with the PREX-2 result of $R^{208}_{\rm skin} = 0.283\pm0.071~\rm{fm}$~\cite{PREX:2021umo}.
The approach proposed in the present work with the expected high precision for the $R^{208}_{\rm skin}$ determination can thus crosscheck the PREX-2 result and test the ab initio prediction on $R^{208}_{\rm skin}$.
In addition,
our approach in principle can be also applied to determine the $R_{\rm skin}$ of other nuclei which are adopted as large-scale detector medium to hunt dark matter and neutrinos, e.g., Xe isotopes in next-generation xenon-based detector~\cite{Aalbers:2022dzr}.

\section{\label{sec:Sum}Conclusions}
We have demonstrated that the neutrinos from a nearby CCSN in our Galaxy can be used to precisely determine the $R_n$ and $R_{\rm skin}$ of Pb via CE$\nu$NS in RES-NOVA. In particular, an ultimate precision of $\sim 0.1\%$ ($\sim 0.006$ fm) for the $R_n$ ($R_{\rm skin}$) of Pb is expected to be achieved in the most optimistic case that the CCSN explosion were to occur at a distance of $\sim 1 \kpc$ from the Earth.
Such a precision of the $R_{\rm skin}$ is significantly higher than that of the existing and planned experiments in terrestrial labs,
and will eventually pin down the density dependence of the symmetry energy and clarify the issue of the tension between CREX and PREX-2 experiments.

\begin{acknowledgments}
The authors thank Ning Zhou for useful discussions.
This work was supported by the National SKA Program of China No. 2020SKA0120300 and the National Natural Science Foundation of China under Grant Nos. 12235010 and 11625521.
\end{acknowledgments}




\bibliography{references}

\end{document}